\documentclass[conference,letterpaper]{IEEEtran}

\usepackage[utf8]{inputenc} 
\usepackage[T1]{fontenc}
\usepackage{url}
\usepackage{ifthen}
\usepackage{cite}
\usepackage[cmex10]{amsmath} 

\usepackage{enumitem}
\usepackage{stfloats}
\usepackage{graphicx} 
\usepackage{amsfonts} 
\usepackage{amsthm} 
\usepackage{amssymb}
\usepackage{arydshln}
\usepackage{stmaryrd}  
\usepackage{bm}
\usepackage{array} 
\usepackage{multirow}
\usepackage{threeparttable} 
\usepackage[ruled,vlined]{algorithm2e}

\usepackage{hyperref}
\usepackage{booktabs}
\usepackage{cleveref}
\usepackage{balance}

\LinesNumbered
\SetAlgoLined

\Crefname{figure}{Fig.}{Figs.}  
\crefname{algocf}{algorithm}{algorithms}
\Crefname{algocf}{Algorithm}{Algorithms}

\DeclareMathOperator*{\argmin}{arg\,min}  
\DeclareMathOperator*{\argmax}{arg\,max}  

\newtheorem{theorem}{Theorem} 
\newtheorem{lemma}{Lemma}    
 
\newtheorem{definition}{Definition}  
\newtheorem{remark}{Remark}
\newtheorem{proposition}{Proposition}

\newcommand{\multiset}[1]{\llbracket #1 \rrbracket}  

\begin{document}
\title{Error-Building Decoding of Linear Block Codes} 

\author{
    Guoda~Qiu$^{*}$, Ling~Liu$^{\dagger}$, Yuejun~Wei$^{\ddagger}$, and Liping~Li$^{*}$ \\
    \small
    $^{*}$State Key Laboratory of Opto-Electronic Information Acquisition and Protection Technology, Anhui University, Hefei 230601, China \\
    $^{\dagger}$Guangzhou Institute of Technology, Xidian University, Guangzhou 510555, China \\
    $^{\ddagger}$School of Computer and Information Engineering, Shanghai Polytechnic University, Shanghai 201209, China \\
    Email: p24301213@stu.ahu.edu.cn; liuling@xidian.edu.cn; yjwei@sspu.edu.cn; liping\_li@ahu.edu.cn
}

\maketitle

\begin{abstract}
This paper proposes a novel maximum-likelihood (ML) soft-decision decoding framework for linear block codes, termed error-building decoding (EBD). The complete decoding process can be performed using only the parity-check matrix, without requiring any other pre-constructed information (such as trellis diagrams or error-pattern lists), and it has the potential to be further customized if specific properties of the code can be exploited. We formally define error-building blocks, and derive a recursive theorem that allows efficient construction of larger locally optimal blocks from smaller ones, thereby effectively searching for the block associated with the most likely error pattern. The EBD framework is further optimized for extended Hamming codes as an example, through offline and online exclusion mechanisms, leading to a substantial complexity reduction without loss of ML performance. Complexity analysis shows that, for extended Hamming codes of lengths 64, 128, and 256, the fully optimized EBD requires approximately an order of magnitude fewer floating-point operations on average than minimum-edge trellis Viterbi decoding at a frame error rate (FER) of $10^{-3}$.
\end{abstract}

\section{Introduction}

Let $\mathcal{C}$ be an $(N, K, d)$ binary linear block code of length $N$, dimension $K$, and minimum Hamming distance $d$. The most straightforward approach to performing maximum-likelihood (ML) soft-decision decoding over an additive white Gaussian noise (AWGN) channel is to compute the Euclidean distance between the received vector and all $2^K$ modulated codewords. 
A more efficient decoding procedure has been developed using trellis-based representations of $\mathcal{C}$ with $\min\{2^K, 2^{N-K}\}$ states\cite{wolf1978trellis}, and various reduced-state trellis constructions have also been proposed to further lower complexity\cite{muder1988trellis,forney1988trellis,kasami1993trellis}.
An alternative approach is syndrome decoding\cite{snyders1989syndrome}, where the most likely error pattern $\bm{e}_{\text{ML}}$ is identified within the coset of $\mathcal{C}$ that contains the hard-decision vector $\bm{b}$, which is obtained by thresholding the received soft message. Decoding then flips the bits of $\bm{b}$ at the positions indicated by the support of $\bm{e}_{\text{ML}}$. Several works\cite{snyders1991rlsd_isit, snyders1991rlsd_tit, snyders2007hamming} have explored eliminating error patterns based on reliability ordering, thereby effectively reducing the search space while still achieving ML soft-decision decoding.
However, these methods typically require auxiliary information about the code in the decoder beyond the parity-check matrix, such as the codeword space, a trellis diagram representation, or a list of candidate error patterns for each possible syndrome. 
Guessing random additive noise decoding (GRAND) offers a more universal decoding framework\cite{duffy2018grand_isit, duffy2019grand_tit}, and soft GRAND (SGRAND) fully utilizes real-valued channel outputs to achieve ML soft-decision decoding\cite{solomon2020sgrand}. Instead of explicitly enumerating possible error patterns, SGRAND guesses the noise pattern from $\mathbb{F}_2^N$ in descending order of noise likelihood. The decoding process stops when the first valid codeword is found. Specifically, a codeword is considered an output if subtracting the guessed noise pattern from $\bm{b}$ yields a vector in $\mathcal{C}$. This guessed noise pattern is precisely $\bm{e}_{\text{ML}}$. 
However, SGRAND's guessing relies somewhat on serial processing, whereas the parallelizable variant ORBGRAND \cite{duffy2022orbgrand} incurs some ML performance loss.

In this paper, we propose a novel ML soft-decision decoding framework, termed error-building decoding (EBD), which utilizes the parity-check matrix and does not rely on predetermined trellis diagrams or (reduced) error-pattern lists. The proposed EBD framework leverages a new concept of error-building blocks to reformulate the problem of finding $\bm{e}_{\text{ML}}$ as the task of identifying the globally optimal error-building block for a given syndrome. The most likely error pattern $\bm{e}_{\text{ML}}$ is associated with this globally optimal error-building block. Consequently, the bit-flip positions for the initial hard-decision vector $\bm{b}$ can be determined, ultimately yielding the ML decoded codeword.

The remainder of this paper is organized as follows. \Cref{sec:ML_criteria} presents the ML decoding criteria for linear block codes. \Cref{sec:EBD} details the general EBD framework, along with the offline and online exclusion mechanisms used for optimization. The customized optimization method for extended Hamming codes and its results are presented in \Cref{sec:decoding_exHam}. Finally, concluding remarks are given in \Cref{sec:conclusion}.

\section{Maximum-Likelihood Decoding Criteria}
\label{sec:ML_criteria}
Throughout this paper, the transmitted codeword $\bm{c}$ (a row vector) is modulated using binary phase shift keying (BPSK) and transmitted over an AWGN channel. The decoder receives the soft message vector  
$
\bm{y} = (\overline{\bm{1}} - 2\bm{c}) + \bm{n},
$
where $\overline{\bm{1}}$ is an all-ones row vector, and $\bm{n} \in \mathbb{R}^N$ is a zero-mean Gaussian noise vector with variance $\sigma^2$. In the log-likelihood ratio (LLR) domain, the reliability vector is given by $\bm{\lambda} = 2\bm{y}/\sigma^2$. 

Let $\bm{H}=[\bm{h}_0,\bm{h}_1,\ldots,\bm{h}_{N-1}]$ be a parity-check matrix of the code $\mathcal{C}$, where each $\bm{h}_i$ is a  column vector of length $N - K$. 
Each valid codeword is denoted by $\bm{c}_j \in \mathcal{C}$, where $j \in \{0, 1, \ldots, 2^K - 1\}$.  
Given the reliability vector $\bm{\lambda}$, \cite{beery1986fht} proved that the maximum-likelihood codeword $\bm{c}_{\text{ML}}$ is obtained as
\begin{equation}
    \bm{c}_{\text{ML}} = \argmax_{\bm{c}_j\in \mathcal{C}}\sum_{k=0}^{N-1}(1-2c_j[k])\lambda[k].
\end{equation}

An equivalent approach to ML decoding is to find the most likely error pattern $\bm{e}_{\text{ML}}$ for the hard-decision vector $ \bm{b} = \text{HD}(\bm{\lambda}) $ and flip $\bm{b}$ accordingly, where $\text{HD}(x) = \mathbb{I}(x<0) $ and \(\mathbb{I}(\cdot)\) denotes the indicator function (1 if the condition holds, 0 otherwise). We refer to $ \bm{e}_j = \bm{c}_j \oplus \bm{b} $ as an error pattern for $\bm{b}$ and its corresponding syndrome $ \bm{s}=\bm{H}\bm{b}^{T} $, and denote the set of all such error patterns by $ \mathcal{P} = \left\{ \bm{e}_0, \bm{e}_1, \ldots, \bm{e}_{2^K-1} \right\} $.  
In this paper, the pattern penalty of $\bm{e}_j$ is defined as $P(\bm{e}_j)=\sum_{k=0}^{N-1}{e}_j[k]\big|\lambda[k]\big|$. 
\Cref{lem:ML} states that ML soft-decision decoding can be achieved by minimizing the pattern penalty among all error patterns for $\bm{s}$, which has been proved in \cite{snyders1989syndrome}.
\begin{lemma}
\label{lem:ML}
    The maximum-likelihood codeword is $\bm{c}_{\text{ML}} = \bm{b} \oplus \bm{e}_{\text{ML}}$, where $\bm{e}_{\text{ML}}$ associated with $\bm{b}$ is given by
    \begin{equation}
    \label{equ:ML}
        \bm{e}_{\text{ML}} = \argmin_{\bm{e}_j\in \mathcal{P}}\sum_{k=0}^{N-1}{e}_j[k]\big|\lambda[k]\big|.
    \end{equation}
\end{lemma}
Each error pattern $\bm{e}_j$ for the syndrome satisfies the relation $\bm{s} = \bm{H}\bm{e}_j^T=\bigoplus_{k=0}^{N-1}e_j[k]\bm{h}_k$. 
Specifically, if certain column vectors in $ \bm{H} $ sum (modulo-2) to $ \bm{s} $, their indices in $ \bm{H} $ form the support set of a valid error pattern. 
We can determine all possible error patterns for $\bm{s}$ according to this rule.

\section{Error-Building Decoding (EBD)}
\label{sec:EBD}
\subsection{Theoretical Foundations of EBD}
\label{subsec:def_and_the}
We introduce the concept of error-building blocks for a vector $\bm{v} \in \mathbb{F}_2^{Q}$, where $Q=N-K$ denotes the number of redundant parity bits. 
An error-building block for $\bm{v} $ is similar to the support set of an error pattern for $\bm{v}$, but is defined as a multiset. Repeated elements are permitted in a multiset, but the order is not considered. To distinguish from the standard set notation $\{ \cdot \}$, a multiset is denoted by $ \multiset{\cdot} $. If a multiset contains no repeated elements, it can be identified with a standard set. The notation $ |\mathcal{A}| $ denotes the size (or cardinality) of a set or multiset $\mathcal{A}$. The symbol $ \cup$ represents the union of two sets, and $ \uplus$ represents the sum (or disjoint union) of two multisets. For instance, $\{1,2,3\} \cup \{2,3,4\} = \{1,2,3,4\}$ and $\multiset{1,1,2} \uplus \multiset{2,2,3} = \multiset{1,1,2,2,2,3}$. 
Additionally, let $\bm{i}$ denote the binary vector corresponding to the integer $i$. 

\begin{definition}
\label{def:ebb}
A multiset $\multiset{a_0,\ldots,a_{t-1} | a_i \in [0,N-1]}$ of size $t$ is referred to as a $t$-error-building block for $\bm{v} =\bigoplus_{i=0}^{t-1}\bm{h}_{a_i}$, where $1\leq t \leq N$. 
Given a vector $\bm{v} \in \mathbb{F}_2^{Q}$, denote  $\bm{\mathcal{B}}_t(\bm{v})$ as the set of all $t$-error-building blocks for $\bm{v}$, and $\mathcal{B}_{t,i}(\bm{v})$ as the $i$-th $t$-error-building block in $\bm{\mathcal{B}}_t(\bm{v})$, i.e., $\bm{\mathcal{B}}_t(\bm{v}) = \{\mathcal{B}_{t,0}(\bm{v}),\dots, \mathcal{B}_{t,T-1}(\bm{v})\}$, 
where $T=|\bm{\mathcal{B}}_t(\bm{v})|$. When no combination of any $ t $ column vectors in the parity-check matrix (repeated columns are permitted) can synthesize $ \bm{v} $, the $t$-error-building block for $\bm{v}$ is $\emptyset$ (empty) and $ \bm{\mathcal{B}}_t(\bm{v}) = \emptyset $. 
Additionally, let $\bm{\mathcal{B}}(\bm{v}) = \bigcup_{t=1}^{N}\bm{\mathcal{B}}_t(\bm{v}) $ denote the overall set of all error-building blocks for $ \bm{v} $.
\end{definition}

To clarify \Cref{def:ebb}, consider a parity-check matrix of the following form: $ \bm{H} = [\bm{h}_0, \bm{h}_1, \bm{h}_2, \bm{h}_3, \bm{h}_4, \bm{h}_5, \bm{h}_6] = [\bm{1}, \bm{2}, \bm{3}, \bm{4}, \bm{5}, \bm{6}, \bm{7}] $, where each column vector $ \bm{h}_i$ in $\bm{H}$ is the 3-bit binary vector corresponding to the decimal number $ i + 1 $. Given a binary vector $\bm{v} = \bm{3} \in \mathbb{F}_2^3$, we have {$\bm{\mathcal{B}}_1(\bm{3})=\{ \multiset{2}\}$}, $\bm{\mathcal{B}}_2(\bm{3})=\{ \multiset{0,1}, \multiset{3,6}, \multiset{4,5} \}$, and $\bm{\mathcal{B}}_3(\bm{3})= \{ \multiset{0,0,2},$ $\multiset{1,1,2},$ $\multiset{2,2,2},$ $\multiset{3,3,2},$ $\multiset{4,4,2},$ $\multiset{5,5,2},$ $[\![6,6,$ $2]\!], \multiset{0,3,5}, \multiset{0,4,6}, \multiset{1,3,4}, \multiset{1,5,6}\}$. 
Similarly, other sets from $\bm{\mathcal{B}}_4(\bm{3})$ to $\bm{\mathcal{B}}_7(\bm{3})$ can be obtained, and finally $ \bm{\mathcal{B}}(\bm{3}) = $ $\bm{\mathcal{B}}_1(\bm{3}) \cup  \bm{\mathcal{B}}_2(\bm{3}) \cup  \cdots \cup \bm{\mathcal{B}}_7(\bm{3}) $.

$\bm{\mathcal{B}}(\bm{v})$ is partitioned into two disjoint subsets: $ \bm{\mathcal{R}}(\bm{v}) $ and $ \bm{\mathcal{U}}(\bm{v}) $, where each block in $ \bm{\mathcal{R}}(\bm{v}) $ contains repeated elements, and each block in $ \bm{\mathcal{U}}(\bm{v}) $ consists of distinct elements only. 
According to \Cref{def:ebb}, each block in $ \bm{\mathcal{U}}(\bm{v}) $ is the support set of an error pattern for $ \bm{v} $.  
Moreover, all support sets of error patterns for $ \bm{v} $ are included in $ \bm{\mathcal{U}}(\bm{v}) $. Note that each block in $\bm{\mathcal{U}}(\bm{v})$ is treated as a standard set here. Therefore, there is a one-to-one mapping between the blocks in $ \bm{\mathcal{U}}(\bm{v}) $ and the error patterns for $\bm{v}$. 
Suppose $ \mathcal{R}'(\bm{v}) $ is a block in $ \bm{\mathcal{R}}(\bm{v}) $.  
After removing all elements that appear an even number of times and retaining only one instance for each element that appears an odd number of times, $ \mathcal{R}'(\bm{v}) $ is transformed into a block in $ \bm{\mathcal{U}}(\bm{v}) $ without repeated elements, denoted by $ \mathcal{U}'(\bm{v}) $.  
We refer to $ \mathcal{U}'(\bm{v}) $ as the reduced form of $ \mathcal{R}'(\bm{v}) $. For example, $\multiset{1,3,4}$ is the reduced form of $\multiset{1,1,1,2,2,3,4}$.

\begin{definition}
\label{def:M}
The block penalty of an empty error-building block is defined as $M(\emptyset) \triangleq +\infty$, and the block penalty of a nonempty $t$-error-building block $\multiset{ a_0,\ldots,a_{t-1}}$
 is defined as
\begin{equation}
    M\big(\multiset{ a_0,\ldots,a_{t-1}}\big) \triangleq \sum_{i=0}^{t-1}\big|\lambda[{a_i}]\big|.
\end{equation}
\end{definition}

\begin{definition}
\label{def:O}
If $\bm{\mathcal{B}}_t(\bm{v}) \neq \emptyset$, the optimal $t$-error-building block for $\bm{v}$ is defined as
\begin{equation}
    \mathcal{O}_t(\bm{v}) \triangleq \argmin_{\mathcal{B}_{t,i}(\bm{v})\in \bm{\mathcal{B}}_t(\bm{v})} M\big(\mathcal{B}_{t,i}(\bm{v})\big).
\end{equation}
If $\bm{\mathcal{B}}_t(\bm{v}) = \emptyset$, let $\mathcal{O}_t(\bm{v}) \triangleq \emptyset$. Any optimal $t$-error-building block is referred to as a locally optimal error-building block. The globally optimal error-building block for $\bm{v}$ is defined as
\begin{equation}
    \mathcal{O}(\bm{v}) \triangleq \argmin_{\mathcal{B}_{t,i}(\bm{v})\in \bm{\mathcal{B}}(\bm{v})} M\big(\mathcal{B}_{t,i}(\bm{v})\big).
\end{equation}
\end{definition}

Note that $\mathcal{O}_t(\bm{v})$ or $\mathcal{O}(\bm{v})$ may not be unique; in this paper, the equals sign used with them denotes a particular solution selected from the set of possible solutions.
Obviously, $\mathcal{O}(\bm{v})$ can be obtained from among $N$ locally optimal error-building blocks for $\bm{v}$. 
The following theorem indicates that this search space can be further reduced. 

\begin{theorem}
\label{the:global_O}
The globally optimal error-building block for $\bm{v}$ can be selected from the first $Q$ locally optimal error-building blocks for $\bm{v}$, i.e., $\mathcal{O}(\bm{v}) = \mathcal{O}_{t^{*}}(\bm{v})$ with $t^{*}\in [1, Q]$, where
\begin{equation}
    t^* = \argmin_{t\in [1,Q] } M\big(\mathcal{O}_{t}(\bm{v})\big).
\end{equation}

\begin{proof}
It suffices to prove that if $\mathcal{O}(\bm{v})$ has a solution of size $> Q$, then it must also have one of size $\leq Q$. 
Assume there exists a $ t' \in [Q+1, N] $ such that $ \mathcal{O}(\bm{v}) = \mathcal{O}_{t'}(\bm{v}) = \multiset{a_0, \dots, a_{t'-1}} $, where $ \bigoplus_{i=0}^{t'-1} \bm{h}_{a_i} = \bm{v}$.
Since the rank of the parity-check matrix $ \bm{H} $ is at most $ Q $, the column vectors $ \bm{h}_{a_0}, \dots, \bm{h}_{a_{t'-1}} $ are linearly dependent over $ \mathbb{F}_2 $, regardless of whether $ \mathcal{O}_{t'}(\bm{v}) $ contains repeated elements. 
This implies that there exist coefficients $ k_0, k_1, \dots, k_{t'-1} \in \{0, 1\} $, not all zero, such that $ \bigoplus_{i=0}^{t'-1} k_i \bm{h}_{a_i} = \bm{0} $.
Combining this equation with the previous equation $ \bigoplus_{i=0}^{t'-1} \bm{h}_{a_i} = \bm{v} $, we obtain $ \bigoplus_{i=0}^{t'-1} (1-k_i) \bm{h}_{a_i} = \bm{v} $.
Since not all $ 1-k_i $~are~equal to one, by \Cref{def:ebb}, there exists a submultiset of $ \multiset{a_0, \dots, a_{t'-1}} $ that can also serve as an error-building block for $ \bm{v} $. Moreover, there necessarily exists such a submultiset of size $ t'' \leq Q $; otherwise, this reduction could continue until one is reached, which is denoted by $\mathcal{B}_{t'',j}(\bm{v})$. 
According to \Cref{def:M}, we have $ M\big( \mathcal{B}_{t'',j}(\bm{v}) \big) \leq M\big( \mathcal{O}_{t'}(\bm{v}) \big) $. 
Combining this inequality with the assumption $ \mathcal{O}(\bm{v}) = \mathcal{O}_{t'}(\bm{v}) $, it follows that $ M\big( \mathcal{B}_{t'',j}(\bm{v}) \big) = M\big( \mathcal{O}_{t'}(\bm{v}) \big)$, and thereby $ \mathcal{O}(\bm{v}) = \mathcal{B}_{t'',j}(\bm{v}) $ is another solution. 
This implies that $\mathcal{O}_{t''}(\bm{v}) = \mathcal{B}_{t'',j}(\bm{v})$ and proves that $ \mathcal{O}(\bm{v}) $ can always be found among all $ \mathcal{O}_t(\bm{v}) $ for $ 1 \leq t \leq Q $.
\end{proof}
\end{theorem}

The following recursive theorem provides an efficient way for constructing larger locally optimal error-building blocks by combining two specific smaller ones. 

\begin{theorem}
\label{the:O}
Given a vector $\bm{v}\in\mathbb{F}_2^Q$, if $M\big(\mathcal{O}_{t_1}(\bm{v}_i)\big) = +\infty$ or $M\big(\mathcal{O}_{t_2}(\bm{v}_i\oplus \bm{v})\big) = +\infty$ holds for all $\bm{v}_i\in\mathbb{F}_2^Q$, then  $ \mathcal{O}_{t_1+t_2}(\bm{v}) = \emptyset$ and $ M\big(\mathcal{O}_{t_1+t_2}(\bm{v})\big) = +\infty$. 
Otherwise, $\mathcal{O}_{t_1+t_2}(\bm{v})$ can be constructed by combining $\mathcal{O}_{t_1}(\bm{v}_b)$ and $\mathcal{O}_{t_2}(\bm{v}_b\oplus \bm{v})$, where $\bm{v}_b$ is a decomposition vector satisfying
\begin{equation}
\label{equ:v_b}
    \bm{v}_b = \argmin_{\bm{v}_i\in\mathbb{F}_2^Q} \big[M\big(\mathcal{O}_{t_1}(\bm{v}_i)\big)+M\big(\mathcal{O}_{t_2}(\bm{v}_i\oplus \bm{v})\big)\big].
\end{equation}
The optimal $(t_1 + t_2)$-error-building block for $\bm{v}$ and corresponding block penalty are given by
\begin{equation}
\label{equ:O}
    \mathcal{O}_{t_1+t_2}(\bm{v}) = \mathcal{O}_{t_1}(\bm{v}_b) \uplus \mathcal{O}_{t_2}(\bm{v}_b\oplus \bm{v}),
\end{equation}
\begin{equation}
\label{equ:Ometric}
    M(\mathcal{O}_{t_1+t_2}(\bm{v})) =  M\big(\mathcal{O}_{t_1}(\bm{v}_b)\big) + M\big(\mathcal{O}_{t_2}(\bm{v}_b \oplus \bm{v})\big).
\end{equation}

\begin{proof}
We first prove the first part.  
Assume there exists a nonempty $ (t_1 + t_2) $-error-building block $ \multiset{a_0, \ldots, a_{t_1 + t_2-1}}$ for $\bm{v}$. Then, the first $ t_1 $ elements form a $ t_1 $-error-building block $ \multiset{a_0, \ldots, a_{t_1-1}} $ for $\bm{v}'=\bigoplus_{i=0}^{t_1-1}\bm{h}_{a_i}$, and the last $ t_2 $ elements form a $ t_2 $-error-building block $ \multiset{a_{t_1}, \ldots, a_{t_1+t_2-1}}$ for $ \bm{v}'' = \bigoplus_{i=t_1}^{t_1+t_2-1} \bm{h}_{a_i}$. Note that $ \bm{v}'' = \bm{v}' \oplus \bm{v} $, which implies that both $ \bm{\mathcal{B}}_{t_1}(\bm{v}') $ and $ \bm{\mathcal{B}}_{t_2}(\bm{v}' \oplus \bm{v}) $ are nonempty.  
According to \Cref{def:M} and \ref{def:O}, this means $ M\big(\mathcal{O}_{t_1}(\bm{v}')\big) \neq +\infty $ and $ M\big(\mathcal{O}_{t_2}(\bm{v}' \oplus \bm{v})\big) \neq +\infty $, which contradicts the given condition.
Therefore, if $M\big(\mathcal{O}_{t_1}(\bm{v}_i)\big) = +\infty$ or $M\big(\mathcal{O}_{t_2}(\bm{v}_i\oplus \bm{v})\big) = +\infty$ holds for all $\bm{v}_i\in\mathbb{F}_2^Q$, then we have $ \bm{\mathcal{B}}_{t_1 + t_2}(\bm{v}) = \emptyset$, and consequently $\mathcal{O}_{t_1 + t_2}(\bm{v}) = \emptyset$ and $M\big(\mathcal{O}_{t_1 + t_2}(\bm{v})\big) = +\infty$. 

We now prove the second part.  
Let $ \mathcal{V} $ denote the set of all vectors $ \bm{v}_i $ that satisfy $ M\big(\mathcal{O}_{t_1}(\bm{v}_i)\big) \neq +\infty $ and $ M\big(\mathcal{O}_{t_2}(\bm{v}_i \oplus \bm{v})\big) \neq +\infty $.
If and only if $ \bm{v}_i \in \mathcal{V} $, both $ \bm{\mathcal{B}}_{t_1}(\bm{v}_i) $ and $ \bm{\mathcal{B}}_{t_2}(\bm{v}_i \oplus \bm{v}) $ are nonempty sets.
According to \Cref{def:ebb}, any  block in nonempty $ \bm{\mathcal{B}}_{t_1}(\bm{v}_i) $ and any block in nonempty $ \bm{\mathcal{B}}_{t_2}(\bm{v}_i \oplus \bm{v}) $ can be combined to form a valid $(t_1 + t_2)$-error-building block for $ \bm{v} $.  
Equation \eqref{equ:v_b} ensures that $ \bm{v}_b \in \mathcal{V} $.  
Therefore, $\mathcal{O}_{t_1}(\bm{v}_b) \uplus \mathcal{O}_{t_2}(\bm{v}_b\oplus \bm{v})$ yields a valid $(t_1 + t_2)$-error-building block in $ \bm{\mathcal{B}}_{t_1+t_2}(\bm{v}) $, and its corresponding block penalty is $M\big(\mathcal{O}_{t_1}(\bm{v}_b)\big) + M\big(\mathcal{O}_{t_2}(\bm{v}_b \oplus \bm{v})\big)$. 

Next, we show that this combination indeed forms an optimal $(t_1+t_2)$-error-building block for $\bm{v}$. The $i$-th $(t_1+t_2)$-error-building block for $\bm{v}$,  $\mathcal{B}_{t_1+t_2, i}(\bm{v})$, is arbitrarily decomposed into two subblocks $\mathcal{B}_{t_1, j_1}(\bm{v}')$ and $\mathcal{B}_{t_2, j_2}(\bm{v}'\oplus \bm{v})$, such that $\mathcal{B}_{t_1+t_2, i}(\bm{v}) = \mathcal{B}_{t_1, j_1}(\bm{v}') \uplus \mathcal{B}_{t_2, j_2}(\bm{v}'\oplus \bm{v})$. Then we have
\begin{align}
\label{equ:Oprove}
    M\big(\mathcal{B}_{t_1+t_2, i}(\bm{v})\big) &=  M\big(\mathcal{B}_{t_1, j_1}(\bm{v}')\big) + M\big(\mathcal{B}_{t_2, j_2}(\bm{v}' \oplus \bm{v})\big) \nonumber \\ 
    &\geq 
    M\big(\mathcal{O}_{t_1}(\bm{v}')\big) + M\big(\mathcal{O}_{t_2}(\bm{v}' \oplus \bm{v})\big) \nonumber \\
    &\geq 
    M\big(\mathcal{O}_{t_1}(\bm{v}_b)\big) + M\big(\mathcal{O}_{t_2}(\bm{v}_b \oplus \bm{v})\big).
\end{align}
The derivation in \eqref{equ:Oprove} indicates that, for any block in $ \bm{\mathcal{B}}_{t_1+t_2}(\bm{v}) $, its corresponding block penalty is no less than that of $\mathcal{O}_{t_1}(\bm{v}_b) \uplus \mathcal{O}_{t_2}(\bm{v}_b \oplus \bm{v})$. 
Therefore, Equation \eqref{equ:O} characterizes a locally optimal block, with the corresponding block penalty specified by \eqref{equ:Ometric}.
\end{proof}
\end{theorem}

The recursive construction of locally optimal error-building blocks can equivalently be implemented in an iterative manner. 
Throughout the rest of this paper, $t_1$ and $t_2$ are used exclusively to denote the decomposition or combination strategy for a $t$-error-building block such that $t=t_1+t_2$.

\begin{theorem}
\label{the:ML}
Given a hard-decision vector $\bm{b}$ and its corresponding syndrome $\bm{s}$, if $\bm{s} = \bm{0}$, the maximum-likelihood codeword is $\bm{c}_{\text{ML}} = \bm{b}$. Otherwise, $\bm{c}_{\text{ML}} = \bm{b} \oplus \bm{e}_{\text{ML}}$, where  $\bm{e}_{\text{ML}}$ is an error pattern associated with $\mathcal{O}(\bm{s})$. Specifically, 
if $ \mathcal{O}(\bm{s}) \in \bm{\mathcal{U}}(\bm{s}) $, 
then it directly provides the support set of $ \bm{e}_{\text{ML}} $; if the solution for $\mathcal{O}(\bm{s})$ found belongs to $ \bm{\mathcal{R}}(\bm{s}) $,  
then its corresponding reduced form provides the support set of $ \bm{e}_{\text{ML}} $. 
\begin{proof}
This theorem can be proved by combining \Cref{lem:ML} with the relationship between error-building blocks and error patterns as previously analyzed. For a detailed proof, please refer to Appendix \ref{app:proof_ML}.    
\end{proof}
\end{theorem}

\subsection{General EBD Framework}
The idea of EBD is to identify error-prone positions in the received vector by leveraging locally optimal error-building blocks. The general decoding framework is as follows: 

\begin{enumerate}[start=0]

\item Compute $\bm{b} = \text{HD}(\bm{\lambda})$ and $\bm{s} = \bm{H}\bm{b}^T$. If $\bm{s} = \bm{0}$, set~$\bm{c}_{\text{ML}} = \bm{b}$ and exit; otherwise, proceed with the following steps.

\item For all $ \bm{v} \in \mathbb{F}_2^Q $, determine $ \bm{\mathcal{B}}_1(\bm{v}) $ using $\bm{H}$ according to \Cref{def:ebb}, and then obtain $ \mathcal{O}_1(\bm{v}) $ and $M\big(\mathcal{O}_{1}(\bm{v})\big)$ with the aid of $\bm{\lambda}$ based on \Cref{def:M,def:O}. 

\item For each integer $t$ from $2$ to $\lceil Q/2 \rceil$, construct $ \mathcal{O}_t(\bm{v}) $ and $M\big(\mathcal{O}_{t}(\bm{v})\big)$ for all $\bm{v}\in\mathbb{F}_2^Q$ via \Cref{the:O} with the setting $(t_1, t_2) = (t - 1, 1)$ or other valid choices.

\item For each integer $t$ from $\lceil Q/2 \rceil + 1$ to $ Q$, construct $ \mathcal{O}_t(\bm{s}) $ and $M\big(\mathcal{O}_{t}(\bm{s})\big)$ via \Cref{the:O} with the setting $(t_1, t_2) = (t - \lceil Q/2 \rceil, \lceil Q/2 \rceil)$ or other valid choices.

\item Apply \Cref{the:ML} to obtain the ML decoded codeword. 

\end{enumerate}

Consider the case $ \bm{s} \neq \bm{0} $ in the following discussion. 
When $ \lceil Q / 2 \rceil + 1 \leq t \leq Q $, the error-building blocks obtained in previous iterations are  sufficient to construct $\mathcal{O}_t(\bm{s}) $; therefore, constructing $ \mathcal{O}_t(\bm{v}) $ for $ \bm{v} \neq \bm{s} $ is unnecessary.
During Steps~2 and 3, if $t$ is even, we may specify $ t_1 = t_2 = t / 2 $ due to the symmetry in the block penalty computation: $M\big(\mathcal{O}_{t/2}(\bm{v}_i)\big) + M\big(\mathcal{O}_{t/2}(\bm{v}_i \oplus \bm{v})\big)
= M\big(\mathcal{O}_{t/2}(\bm{v}_i \oplus \bm{v})\big) + M\big(\mathcal{O}_{t/2}((\bm{v}_i \oplus \bm{v}) \oplus \bm{v})\big)$, 
which implies that, for a fixed nonzero vector $\bm{v}$, searching over all $\bm{v}_i \in \mathbb{F}_2^Q$ in \eqref{equ:v_b} can be halved by enumerating the $2^{Q-1}$ unordered vector pairs of the form $\{\bm{v}_i,\, \bm{v}_i \oplus \bm{v}\}$.

Suppose there are no repeated columns in $\bm{H}$. 
The general EBD framework requires iteratively constructing a total of  $ \lceil Q / 2-1 \rceil \times 2^Q + \lfloor Q / 2 \rfloor \times 1 $ locally optimal error-building blocks of size greater than 1.  
For each block, the computation involves at most $ 2^Q $ additions and $ 2^Q - 1 $ comparisons according to \eqref{equ:v_b}. Additionally, the final selection of $\mathcal{O}(\bm{s})$ needs $Q-1$ comparisons. 
As a result, the upper bound on the number of floating-point operations is $ \lceil Q / 2-1 \rceil \times 2^{2Q+1} + \lfloor Q / 2 \rfloor \times 2^{Q+1} $, making it suitable for decoding codes with small $ Q $. 
In practice, symmetry can be exploited (as discussed above), and additions and comparisons involving $ +\infty $ can be safely ignored. Consequently, the total complexity is reduced.

\begin{figure}[t]
    \centering
    \includegraphics[width=1\linewidth]{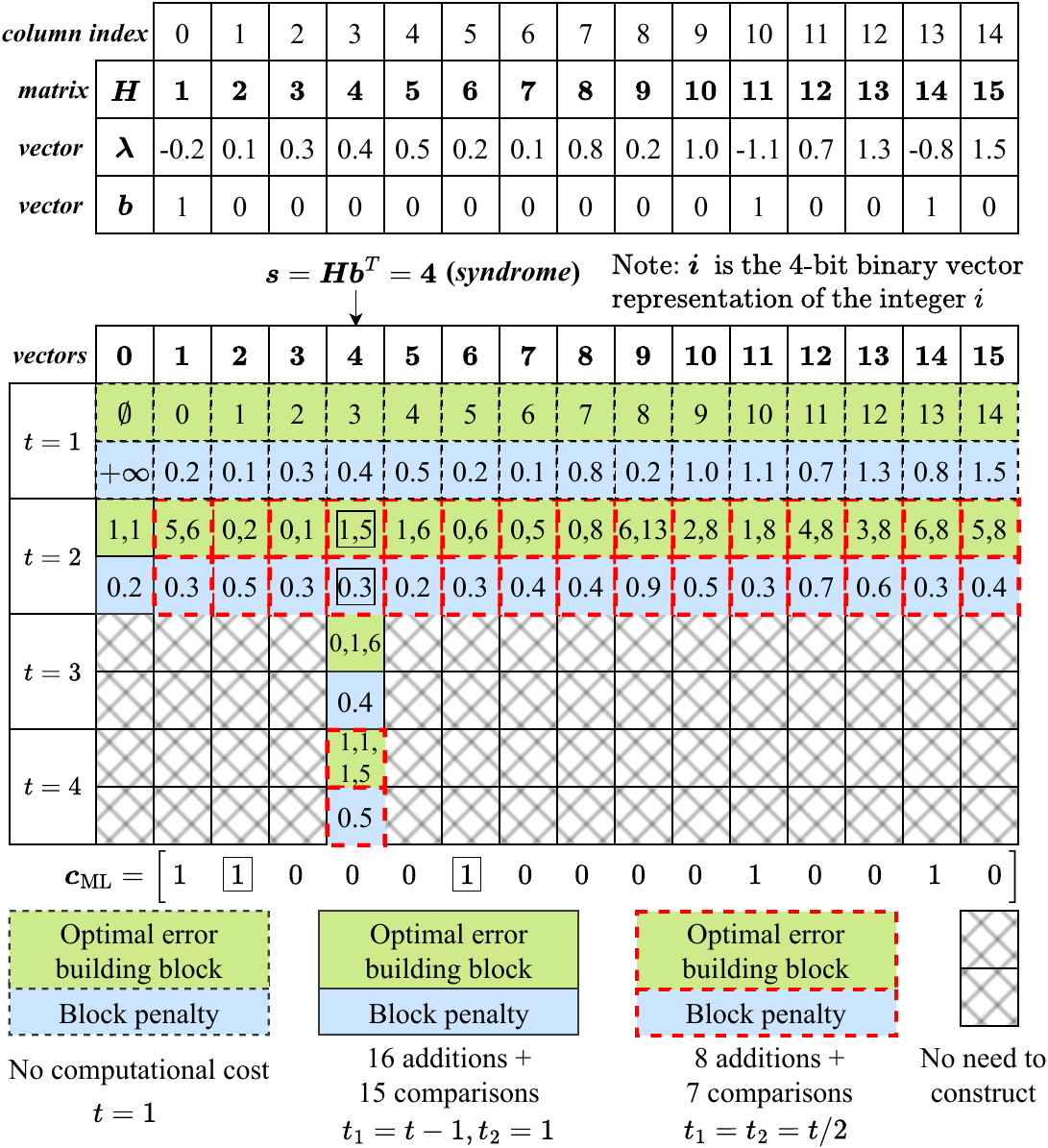}
    \caption{An example of decoding the $(15,11,3)$ Hamming code using EBD.}
    \label{fig:ML_hamming}
\end{figure}

Take the $(15,11,3)$ Hamming code as an example. Its parity-check matrix is formed by listing all nonzero vectors in $\mathbb{F}_2^4$ as columns, i.e., $\bm{H} = [\bm{1}, \bm{2}, \ldots, \bm{15}]$, where $\bm{h}_i$ denotes the 4-bit binary representation of the integer $i+1$ for $i \in [0,14]$.
Assume the reliability vector is $ \bm{\lambda} = [-0.2, 0.1, \ldots, 1.5] $ (refer to \Cref{fig:ML_hamming} for details). 
The resulting hard-decision vector is  
$ \bm{b} = [1, 0, \ldots, 0]$, and the syndrome is $ \bm{s} = \bm{4} $.
First, optimal 1-error-building blocks for all $ \bm{v} \in \mathbb{F}_2^4 $ can be directly obtained. 
For example, $ \mathcal{O}_1(\bm{1}) = \multiset{0} $ since only $ \bm{h}_0 = \bm{1} $, and its block penalty is $ M\big(\mathcal{O}_1(\bm{1})\big)  = \big|\lambda[0]\big| = 0.2 $.
Moreover, $ \mathcal{O}_1(\bm{0}) = \emptyset $ and $ M\big(\mathcal{O}_1(\bm{0})\big) = +\infty $, because~no column in $ \bm{H} $ corresponds to the zero vector $\bm{0}$.
Next, by setting $ t_1 = t_2 = 1 $, we construct optimal 2-error-building blocks for all $ \bm{v} \in \mathbb{F}_2^4 $.  
In particular, when $ \bm{v} \neq \bm{0} $, the computation of \eqref{equ:v_b} can be halved by grouping vector pairs.
For example, the 8 vector pairs associated with $ \bm{1} $ are $ \{\bm{0}, \bm{1}\}, \{\bm{2}, \bm{3}\}, \ldots, \{\bm{14}, \bm{15}\}$.  
Therefore, only 8 additions and 7 comparisons among $ \{
M\big(\mathcal{O}_1(\bm{0})\big) \mkern -1mu + \mkern -1mu M\big(\mathcal{O}_1(\bm{1})\big), M\big(\mathcal{O}_1(\bm{2})\big) \mkern -1mu + \mkern -1mu M\big(\mathcal{O}_1(\bm{3})\big), \ldots, M\big(\mathcal{O}_1(\bm{14})\big) \mkern -1mu + \mkern -1mu M\big(\mathcal{O}_1(\bm{15})\big) \} $  
are required to construct $ \mathcal{O}_2(\bm{1}) $.
Then, the construction of $ \mathcal{O}_3(\bm{4}) $ is achieved by setting $t_1=2$ and $t_2=1$, while $ \mathcal{O}_4(\bm{4}) $ is built~using $ t_1 = t_2 = 2 $ to reduce the computational cost by half. 
\Cref{fig:ML_hamming} shows all  error-building blocks generated during the decoding process for this example.
Finally, $ \mathcal{O}(\bm{4}) = \mathcal{O}_2(\bm{4}) = \multiset{1, 5} $, which implies that flipping bits at positions 1 and 5 in $ \bm{b} $ yields the maximum-likelihood codeword.
Note that all indices are zero-based.
When additions and comparisons involving $+\infty$ are considered, as in \Cref{fig:ML_hamming}, the total number of operations is 305. 
However, terms involving $+\infty$ can be safely ignored, reducing the actual count to 271. 
For comparison, the approach in \cite{beery1986fht} requires 254 operations to decode the $(15, 11, 3)$ Hamming code, while \cite{snyders1989syndrome} can perform decoding with at most 83 operations, and the method used in \cite{wolf1978trellis} requires 409 operations.

\subsection{Optimization via Offline and Online Exclusion Mechanisms}
\label{subsec:mechanisms}
The offline exclusion mechanism fixes certain blocks as empty (i.e., assigning each block a penalty of $+\infty$) throughout the entire decoding process, while the online exclusion mechanism dynamically renders certain computed blocks empty. 
This optimized decoding adopts the same block-combination logic as the general EBD and reduces practical computational costs by yielding more empty blocks; however, the resulting constructed blocks may no longer be locally optimal for their respective vectors.
Consequently, we replace the construction of any $\mathcal{O}_t(\bm{v})$ requiring evaluation with that of $\mathcal{Z}_t(\bm{v})$, where $\mathcal{Z}_t(\bm{v})$ denotes the target $t$-error-building block for $\bm{v}$.
Let $\mathcal{Z}^{t}(\bm{s})$ denote the block with the minimum penalty among $\{\mathcal{Z}_1(\bm{s}),\mathcal{Z}_2(\bm{s}), \ldots, \mathcal{Z}_{t}(\bm{s})\}$. 
Now, the decoding goal becomes selecting $\mathcal{Z}^{Q}(\bm{s})$. 
The decoding process is detailed as follows.

For any $ t $ and $ \bm{v} $, if $ \mathcal{O}_t(\bm{v}) $ can be determined prior to decoding to be empty or to contain repeated elements, then the corresponding target block $ \mathcal{Z}_t(\bm{v}) $ is deemed offline-excludable and is fixed to $ \emptyset $.
If $\mathcal{Z}_1(\bm{v})$ is not offline-excludable, initialize $\mathcal{Z}_1(\bm{v}) = \mathcal{O}_1(\bm{v})$. 
During the subsequent iterative construction process, the non-offline-excludable block $\mathcal{Z}_{t}(\bm{v})$ is constructed by $\mathcal{Z}_{t}(\bm{v}) = \mathcal{Z}_{t_1}(\bm{v}_b) \uplus \mathcal{Z}_{t_2}(\bm{v}_b \oplus \bm{v})$, where $t=t_1+t_2$ and $\bm{v}_b$ is one of the $\bm{v}_i$ to minimize $M\big(\mathcal{Z}_{t_1}(\bm{v}_i)\big)+M\big(\mathcal{Z}_{t_2}(\bm{v}_i\oplus \bm{v})\big)$.  
If the online exclusion mechanism is also integrated, it further excludes certain computed blocks that do not contribute to ML decoding by setting them to empty. 
Specifically, before constructing $\mathcal{Z}_t(\bm{v})$ with $t = t_1 + t_2$, we first need to obtain $\mathcal{D}_{t_j}^{\,t-1}(\bm{v}_i)$ for any $\bm{v}_i\in \mathbb{F}_2^Q$, where $t_j\in\{t_1,t_2\}$, and 
\begin{equation}
\label{equ:online_D}
    \mathcal{D}_{t_j}^{t-1}(\bm{v}_i) \!= \!
    \begin{cases}
    \mathcal{Z}_{t_j}(\bm{v}_i), 
    \!\!\!\! & \text{if } M\big(\mathcal{Z}_{t_j}(\bm{v}_i)\big) \!<\! M\big(\mathcal{Z}^{t-1}(\bm{s})\big), \\
    \emptyset,  \!\!\!\! & \text{if } M\big(\mathcal{Z}_{t_j}(\bm{v}_i)\big) \!\geq\! M\big(\mathcal{Z}^{t-1}(\bm{s})\big).
    \end{cases}
\end{equation}
If $\mathcal{D}_{t_1}^{t-1}(\bm{v}_i)$ or $\mathcal{D}_{t_2}^{t-1}(\bm{v}_i\oplus \bm{v})$ is $\emptyset$ for all $\bm{v}_i\in\mathbb{F}_2^Q$, $\mathcal{Z}_{t}(\bm{v}) = \emptyset$; otherwise, 
$ \mathcal{Z}_{t}(\bm{v}) = \mathcal{D}_{t_1}^{t-1}(\bm{v}_b) \uplus \mathcal{D}_{t_2}^{t-1}(\bm{v}_b\oplus \bm{v}) $, 
where $\bm{v}_b$ is one of the $\bm{v}_i$ to minimize $M\big(\mathcal{D}_{t_1}^{t-1}(\bm{v}_i)\big)+M\big(\mathcal{D}_{t_2}^{t-1}(\bm{v}_i\oplus \bm{v})\big)$.

Appendix \ref{app:proof_onoff_ML} rigorously proves that $\mathcal{O}(\bm{s}) = \mathcal{Z}^{Q}(\bm{s})$. 
Notably, offline-excludable blocks are identified via the algebraic properties of different codes and can be used to optimize the iterative construction strategy, as illustrated in \Cref{subsec:off_optimize_exHam_ML} using extended Hamming codes as an example.

\section{Decoding of Extended Hamming Codes}
\label{sec:decoding_exHam}
\subsection{Customized Offline Optimization}
\label{subsec:off_optimize_exHam_ML}
An extended Hamming code is constructed by taking a standard Hamming code and appending an overall parity bit\cite{macwilliams1977theory}. We exploit the property that its parity-check matrix $\bm{H}$ contains an all-ones row to optimize the EBD framework.
\begin{equation}
\label{equ:H}
    \bm{H} = 
    \left[
    \begin{array}{ccccccccc}
    1 & 1 & 1 & 1 & \cdots & 1 & 1 & 1 & 1  \\
    0 & 1 & 0 & 1 & \cdots& 0 & 1 & 0 & 1  \\
    0 & 0 & 1 & 1 & \cdots& 0 & 0 & 1 & 1  \\
    \vdots & \vdots & \vdots & \vdots & \vdots& \vdots & \vdots & \vdots &  \vdots \\
    0 & 0 & 0 & 0 & \cdots& 1 & 1 & 1 & 1  \\
    \end{array}
    \right]. 
\end{equation}

We consider the form of $\bm{H}$ shown in \eqref{equ:H}, where for any $ i \in [0, N - 1] $, $ {h}_i[0] = 1 $. 
Since the modulo-2 sum of any odd number of columns of $\bm{H}$ yields a vector whose first bit (index 0) is one, it follows that if $v[0] = 0$, then for any~odd $t$, $\mathcal{O}_t(\bm{v})$ is empty and thus $\mathcal{Z}_t(\bm{v})$ is offline-excludable.
Similarly, if $ v[0] = 1 $, $ \mathcal{Z}_t(\bm{v}) $ is offline-excludable for any even $ t $. 
Since even-sized locally optimal blocks for $\bm{v}=\bm{0}$ must consist of repeated elements with the least reliability, their corresponding target blocks are also offline-excludable.
Denote the set of all $ \bm{v} \in \mathbb{F}_2^Q $ such that $ \bm{v} \neq \bm{0} $ and $ v[0] = 0 $ by $ {\mathcal{W}} $, and the set of all $ \bm{v} \in \mathbb{F}_2^Q $ with $ v[0] = 1 $ by $ {\mathcal{Y}} $.  
It suffices to construct target error-building blocks of even size for $ \bm{v} \in \mathcal{W} $ and of odd size for $ \bm{v} \in \mathcal{Y} $.  
Moreover, the iteration step size of the general EBD framework can be relaxed from 1 to 2 in certain cases, provided that all even-sized or odd-sized $\mathcal{Z}_t(\bm{s})$ with $t\in[1,Q]$ can be obtained when $\bm{s} \in \mathcal{W}$ or $\bm{s} \in \mathcal{Y}$, respectively. 
More details are provided in Appendix \ref{app:offline_details}.

\subsection{Results}
\label{subsec:results}
\begin{figure}[t]
    \centering
    \includegraphics[width=0.98\linewidth]{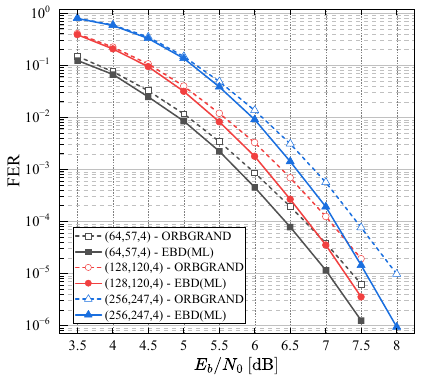}
    \caption{FER of extended Hamming codes of lengths 64, 128, and 256 under EBD (general or optimized framework) and ORBGRAND (up to $10^6$ guesses).}
    \label{fig:FER_exham}
\end{figure}

\begin{table}[htbp]
\centering
\begin{threeparttable}
\caption{Computational Complexity of Decoding Extended Hamming Codes when the Syndrome $\bm{s}\neq\bm{0}$}
\label{tab:Complexity_exHam}
\setlength{\tabcolsep}{5.25pt}  
\begin{tabular}{c|cccccc}

\toprule

\multirow{2}{*}{Code} & Trellis & \multicolumn{2}{c}{EBD-OffOpt ($s[0]$)\tnote{†}}  & \multicolumn{3}{c}{EBD-FullOpt (FER)\tnote{†}} \\

&  \cite{tapp1999trellis_exHam}   &    $0$   &    $1$       & $10^{-2}$     & $10^{-3}$     & $10^{-4}$ \\

\midrule

$(64,57,4)$      & 7531  & 7937  & 16065  & 1397  & 839    & 566\\
$(128,120,4)$  & 31435  & 32383  & 64897  &  5643 & 3231    & 1957\\
$(256,247,4)$    & 128395  & 130303  & 261885  & 23952  & 13213    & 7664\\

\bottomrule

\end{tabular}

\begin{tablenotes}
    \item[†] Different values in parentheses are provided in the following line, and the resulting computational cost depends on the specific choice.
\end{tablenotes}

\end{threeparttable}
\end{table}
EBD with only the offline exclusion mechanism is referred to as EBD-OffOpt, while the version fully optimized through both the offline and online exclusion mechanisms is called EBD-FullOpt.
The optimized iterative construction strategy adopted in our experiment and further details regarding online exclusion are provided in Appendices \ref{app:offline_details} and \ref{app:online_details}, respectively.
\Cref{fig:FER_exham}~shows the frame error rate (FER) of extended Hamming codes under EBD and ORBGRAND \cite{duffy2022orbgrand} versus the ratio of energy per bit to noise power spectral density ($E_b/N_0$). Since both the general and optimized EBD frameworks are ML decoding approaches, their curves coincide and outperform ORBGRAND.
\Cref{tab:Complexity_exHam} presents the computational complexity of EBD-OffOpt and EBD-FullOpt in terms of floating-point operations (counting additions and comparisons that do not involve $+\infty$), with  minimum-edge trellis Viterbi ML decoding as a baseline, which exhibits a constant cost for a given code as shown in \cite{tapp1999trellis_exHam}.  
The complexity of EBD-OffOpt depends on $ s[0] $, while that of EBD-FullOpt is averaged over cases where $\bm{s} \neq \bm{0}$, reflecting the average number of floating-point operations under different FER targets.
Although EBD-OffOpt has a higher cost than trellis decoding, the average cost of EBD-FullOpt is lower than that of trellis decoding after incorporating the online exclusion mechanism, and it decreases as the $E_b/N_0$ increases. 
For the extended Hamming codes $(64,57,4)$, $(128,120,4)$ and $(256,247,4)$ at FER = $10^{-3}$, the average computational cost of EBD-FullOpt is about 11.1\%, 10.3\%, and 10.3\% of that of trellis decoding, respectively.

\section{Conclusion}
\label{sec:conclusion}
EBD is applicable to low-redundancy linear block codes for codeword recovery, provided that the parity-check matrix is known.
This framework reformulates ML decoding as the task of finding the globally optimal error-building block for a given syndrome, where this block corresponds to the most likely error pattern and is obtained from a collection of locally optimal blocks of limited size.  
The theorem on recursive construction enables an efficient iterative construction of these locally optimal blocks.  
For extended Hamming codes, the special structure of the parity-check matrix provides an additional complexity-reduction method via the offline exclusion, which can be combined with the dynamic online exclusion to discard non-contributing blocks early in the decoding procedure.
The fully optimized EBD for extended Hamming codes significantly reduces the average computational cost, particularly at high $E_b/N_0$.
This error-building approach has good generality, and the complexity-reduction techniques can be potentially extended to other algebraic codes.

\newpage


\bibliographystyle{IEEEtran}
\bibliography{references}

\clearpage

\appendices

\section{\texorpdfstring{Proof of \Cref{the:ML}}{Proof of Theorem 3}}
\label{app:proof_ML}
We first prove the first part. Given a syndrome $\bm{s}$, each error pattern $\bm{e}_j$ satisfies the relation $\bm{s} = \bm{H}\bm{e}_j^T=\bigoplus_{k=0}^{N-1}e_j[k]\bm{h}_k$. Therefore, if $\bm{s} = \bm{0}$, the all-zero row vector $\overline{\bm{0}}$ of length $N$ is a valid error pattern and achieves the minimum pattern penalty (which equals 0) among all error patterns for $\bm{s}$. According to \Cref{lem:ML}, we have $\bm{e}_{\text{ML}} = \overline{\bm{0}}$, and thus $\bm{c}_{\text{ML}} = \bm{b} \oplus \bm{e}_{\text{ML}} = \bm{b}$.

We now prove the second part. Based on \Cref{lem:ML} and the one-to-one mapping between error patterns and error-building blocks in $\bm{\mathcal{U}}(\bm{s})$, 
we know that the block in $ \bm{\mathcal{U}}(\bm{s}) $ with the minimum block penalty provides the support set of $\bm{e}_{\text{ML}}$. 
Next, we only need to prove that finding the error-building block with the minimum block penalty over $ \bm{\mathcal{U}}(\bm{s}) $ is equivalent to doing so over $ \bm{\mathcal{B}}(\bm{s}) $, i.e., finding $\mathcal{O}(\bm{s})$, which may have multiple solutions.  
If all solutions lie in $ \bm{\mathcal{U}}(\bm{s}) $, then the one selected as $ \mathcal{O}(\bm{s}) $ is guaranteed to have the minimum block penalty in $ \bm{\mathcal{U}}(\bm{s}) $. 
We now consider the case where one of these solutions lies in $ \bm{\mathcal{R}}(\bm{s}) $, denoted by $ \mathcal{O}(\bm{s}) = \mathcal{R}'(\bm{s}) $,  
where $ \mathcal{R}'(\bm{s}) $ is a block in $ \bm{\mathcal{R}}(\bm{s}) $. 
By carefully handling the repeated elements, the reduced form $ \mathcal{U}'(\bm{s}) $ of $ \mathcal{R}'(\bm{s}) $, satisfies $ M\big( \mathcal{U}'(\bm{s}) \big) \leq M\big( \mathcal{R}'(\bm{s}) \big)$. 
Under the assumption that $ \mathcal{R}'(\bm{s}) $ achieves the minimum block penalty over $ \bm{\mathcal{B}}(\bm{s}) $,  
we must have $ M\big( \mathcal{U}'(\bm{s}) \big) = M\big( \mathcal{R}'(\bm{s}) \big)$, and $ \mathcal{O}(\bm{s}) = \mathcal{U}'(\bm{s}) $ is another valid solution.
Since $ \mathcal{U}'(\bm{s}) $ is the block with the minimum block penalty in $ \bm{\mathcal{B}}(\bm{s}) $, it is also the minimum penalty block in $ \bm{\mathcal{U}}(\bm{s}) $, and it provides the support set of $ \bm{e}_{\text{ML}} $.
This shows that even if $ \mathcal{O}(\bm{s}) $ is found in $ \bm{\mathcal{R}}(\bm{s}) $, its reduced form still provides the error-building block in $ \bm{\mathcal{U}}(\bm{s}) $ with the minimum block penalty.

\section{\texorpdfstring{Proof of $\mathcal{O}(\bm{s}) = \mathcal{Z}^Q(\bm{s})$}{Proof of O(s) = Z\^Q(s)}}
\label{app:proof_onoff_ML}
We have noted in \Cref{sec:EBD} that the equals sign used with $\mathcal{O}(\bm{s})$ denotes a particular solution. Here, $\mathcal{Z}^Q(\bm{s})$ may also not be unique. We only need to prove that among $\{\mathcal{Z}_1(\bm{s}),\mathcal{Z}_2(\bm{s}), \ldots, \mathcal{Z}_{Q}(\bm{s})\}$, there exists one block with the minimum penalty that forms a globally optimal error-building block for $\bm{s}$. For ease of the description, we first define the concept of the minimal flipping block and then use this concept to derive \Cref{pro:O_and_Z} (the core of our proof). \Cref{pro:O_and_Z} shows that the optimized error-building decoding, which integrates offline exclusion, online exclusion, or both, guarantees that certain locally optimal blocks can always be obtained.

\begin{definition}
\label{def:minimal_flipping_block}
Given $\bm{s} \neq \bm{0}$, the block $\mathcal{O}_t(\bm{s})$ is called a minimal flipping block if the following two conditions hold:
\begin{enumerate}[label=(\arabic*)]
    \item $\mathcal{O}_t(\bm{s})$ is a globally optimal error-building block for $\bm{s}$;
    \item there exists no $t' < t$ such that $\mathcal{O}_{t'}(\bm{s})$ is a globally optimal error-building block for $\bm{s}$.
\end{enumerate}
\end{definition}

\begin{proposition}
\label{pro:O_and_Z}
Given a minimal flipping block $ \mathcal{O}_t(\bm{s}) $, let $\mathcal{X}_{t',i'}$ denote a subblock of $\mathcal{O}_t(\bm{s})$ of size $t'$, where $t'\in [1,t]$ and the index $i' \in \big[0, \binom{t}{t'} - 1\big]$ is used to distinguish distinct subblocks of the same size $t'$. 
Define the vector $\bm{v}_{\mathcal{X}_{t',i'}} = \bigoplus_{j \in \mathcal{X}_{t',i'}} \bm{h}_j$. 
Then, the following equation holds:  
\begin{equation}
\label{equ:M_Z=M_O}
    M\big( \mathcal{Z}_{t'}(\bm{v}_{\mathcal{X}_{t',i'}}) \big) = M\big( \mathcal{O}_{t'}(\bm{v}_{\mathcal{X}_{t',i'}}) \big).    
\end{equation}
\begin{proof}
Note that in \Cref{pro:O_and_Z}, $ t $ is a known constant such that $ \mathcal{O}_t(\bm{s}) $ is a minimal flipping block, which is an external parameter dependent on the received soft message vector, and we only need to prove that \eqref{equ:M_Z=M_O} holds for all $ t' \in [1, t] $.
In this proof, we consider only the case where both the offline and online exclusion mechanisms are integrated; similar derivations apply to scenarios with only one mechanism.

For the convenience of the subsequent proof, we first establish the following  claims. Claims 1 and 2 are based on the premise that $\mathcal{O}_t(\bm{s})$ is a minimal flipping block, while Claim 3 is an intrinsic property of the error-building decoding process.
\begin{enumerate}[label=\textbf{Claim \arabic*}, leftmargin=*]
    \item If $t\geq 2$, $M\big(\mathcal{O}_t({\bm{s}})\big) < M\big( \mathcal{O}^{t-1}(\bm{s}) \big)$, where $\mathcal{O}^{t-1}(\bm{s})$ is the block with the minimum penalty among the first $t-1$ locally optimal blocks for $\bm{s}$. 

    \item 
    $\mathcal{O}_{t'}(\bm{v}_{\mathcal{X}_{t',i'}}) = \mathcal{X}_{t',i'}$,
    where $1 \leq t'\leq t$. 

    \item 
    $M\big(\mathcal{Z}_{\bar{t}}(\bm{v})\big) \geq M\big(\mathcal{O}_{\bar{t}}(\bm{v})\big)$, where $\bar{t} \geq 1$ and $\bm{v}\in\mathbb{F}_2^Q$. 
\end{enumerate}

We formally proceed by induction on $ t' $ over the range $[1, t]$. For the base case $t' = 1$, without loss of generality, let $\mathcal{X}_{1,0} = \multiset{a}$, where $a$ is an element of $\mathcal{O}_t(\bm{s})$. By Claim 2, we have $\mathcal{O}_{1}(\bm{v}_{\mathcal{X}_{1,0}}) = \multiset{a}$. Hence, the corresponding target $1$-error-building block is not offline-excludable and satisfies $\mathcal{Z}_1(\bm{v}_{\mathcal{X}_{1,0}}) = \mathcal{O}_{1}(\bm{v}_{\mathcal{X}_{1,0}})$. Similar conclusions hold for any $\mathcal{X}_{1,i'}$ with $i'\in [0,t-1]$, meaning that the base case $t'=1$ yields $M\big( \mathcal{Z}_{t'}(\bm{v}_{\mathcal{X}_{t',i'}}) \big) = M\big( \mathcal{O}_{t'}(\bm{v}_{\mathcal{X}_{t',i'}}) \big)$. Note that if $ t = 1 $, the base case already ensures that \Cref{pro:O_and_Z} holds.

We now carry out the inductive step by considering $ t' \in [2, t] $, and note that this inductive step is needed only when $ t \geq 2 $. 
The induction hypothesis assumes that for all $t'' \in[1, t' - 1]$, the equality $M\big( \mathcal{Z}_{t''}(\bm{v}_{\mathcal{X}_{t'',i''}}) \big) = M\big( \mathcal{O}_{t''}(\bm{v}_{\mathcal{X}_{t'',i''}}) \big)$ holds for $i'' \in \big[0,\, \binom{t}{t''} - 1\big]$. 
For a given $\mathcal{X}_{t',i'}$, $\mathcal{Z}_{t'}(\bm{v}_{\mathcal{X}_{t',i'}})$ is constructed by combining two specific smaller error-building blocks with the setting $t' = t'_1 + t'_2$, where $ t'_1,t'_2 \in [1,t' - 1]$. There must exist two disjoint subblocks of $\mathcal{X}_{t',i'}$, denoted by $\mathcal{X}_{t'_1,j}$ and $\mathcal{X}_{t'_2,k}$, such that $\mathcal{X}_{t'_1,j} \uplus \mathcal{X}_{t'_2,k} = \mathcal{X}_{t',i'}$ (i.e., $\mathcal{X}_{t'_1,j} \cap \mathcal{X}_{t'_2,k} = \emptyset$). 
By the induction hypothesis, we have $ M\big( \mathcal{Z}_{t'_1}(\bm{v}_{\mathcal{X}_{t'_1,j}}) \big) = M\big( \mathcal{O}_{t'_1}(\bm{v}_{\mathcal{X}_{t'_1,j}}) \big) $ and $ M\big( \mathcal{Z}_{t'_2}(\bm{v}_{\mathcal{X}_{t'_2,k}}) \big) = M\big( \mathcal{O}_{t'_2}(\bm{v}_{\mathcal{X}_{t'_2,k}}) \big) $. 
Next, based on the containment relationships among these subblocks and Claim 2, we have
\begin{equation}
    M\big( \mathcal{O}_{t'_1}(\bm{v}_{\mathcal{X}_{t'_1,j}}) \big)
    \leq  M\big( \mathcal{O}_{t'}(\bm{v}_{\mathcal{X}_{t',i'}}) \big) 
    \leq  M\big( \mathcal{O}_{t}(\bm{s}) \big).
\end{equation}  
By Claims 1 and 3, we have
\begin{equation}
    M\big( \mathcal{O}_{t}(\bm{s}) \big) < 
    M\big( \mathcal{O}^{t'-1}(\bm{s}) \big)
    \leq 
    M\big( \mathcal{Z}^{t'-1}(\bm{s}) \big).
\end{equation}  
Therefore, $M\big( \mathcal{Z}_{t'_1}(\bm{v}_{\mathcal{X}_{t'_1,j}}) \big) < M\big( \mathcal{Z}^{t'-1}(\bm{s}) \big)$ holds. 
Based on  \eqref{equ:online_D}, $ \mathcal{D}_{t'_1}^{\,t'-1}(\bm{v}_{\mathcal{X}_{t'_1,j}}) = \mathcal{Z}_{t'_1}(\bm{v}_{\mathcal{X}_{t'_1,j}}) $.
Similarly, we obtain $ \mathcal{D}_{t'_2}^{\,t'-1}(\bm{v}_{\mathcal{X}_{t'_2,k}}) = \mathcal{Z}_{t'_2}(\bm{v}_{\mathcal{X}_{t'_2,k}}) $.
Then, according to the construction of $\mathcal{Z}_{t'}(\bm{v}_{\mathcal{X}_{t',i'}})$, we have
\begin{align}
\label{equ:M_Z_leq_M_O}
M\big( \mathcal{Z}_{t'}(\bm{v}_{\mathcal{X}_{t',i'}}) \big) 
&\le M\big( \mathcal{D}_{t'_1}^{t'-1}(\bm{v}_{\mathcal{X}_{t'_1,j}}) \big) + M\big( \mathcal{D}_{t'_2}^{t'-1}(\bm{v}_{\mathcal{X}_{t'_2,k}}) \big) \notag\\ 
&= M\big( \mathcal{Z}_{t'_1}(\bm{v}_{\mathcal{X}_{t'_1,j}}) \big) + M\big( \mathcal{Z}_{t'_2}(\bm{v}_{\mathcal{X}_{t'_2,k}}) \big) \notag\\ 
&= M\big( \mathcal{O}_{t'_1}(\bm{v}_{\mathcal{X}_{t'_1,j}}) \big) + M\big( \mathcal{O}_{t'_2}(\bm{v}_{\mathcal{X}_{t'_2,k}}) \big) \notag\\
&= M\big( \mathcal{O}_{t'}(\bm{v}_{\mathcal{X}_{t',i'}}) \big). 
\end{align}
By Claim 3, the  decoding process inherently ensures 
\begin{equation}
\label{equ:M_Z_geq_M_O}
M\big( \mathcal{Z}_{t'}(\bm{v}_{\mathcal{X}_{t',i'}}) \big) \ge M\big( \mathcal{O}_{t'}(\bm{v}_{\mathcal{X}_{t',i'}}) \big).
\end{equation}
Finally, combining \eqref{equ:M_Z_leq_M_O} and \eqref{equ:M_Z_geq_M_O}, we obtain 
\begin{equation}
M\big( \mathcal{Z}_{t'}(\bm{v}_{\mathcal{X}_{t',i'}}) \big) = M\big( \mathcal{O}_{t'}(\bm{v}_{\mathcal{X}_{t',i'}}) \big).
\end{equation}

Since the basis step has been verified and the inductive step has been established under the strong induction hypothesis, it follows that \eqref{equ:M_Z=M_O} holds for all $t'\in[1,t]$.
\end{proof}

\end{proposition}

If $\mathcal{O}_t(\bm{s})$ is a minimal flipping block, it follows readily from \Cref{the:global_O} that $t\in[1,Q]$. By setting $t'=t$ in \Cref{pro:O_and_Z}, we obtain $M\big( \mathcal{Z}_{t}(\bm{s})\big) = M\big( \mathcal{O}_{t}(\bm{s})\big)$. Clearly, $\mathcal{Z}_t(\bm{s})$ is a globally optimal error-building block for $\bm{s}$ and achieves the minimum penalty among $\{\mathcal{Z}_1(\bm{s}),\mathcal{Z}_2(\bm{s}) \ldots, \mathcal{Z}_{Q}(\bm{s})\}$.
Therefore, we conclude that
$\mathcal{O}(\bm{s}) = \mathcal{Z}^Q(\bm{s})$.

\section{Optimized Iterative Construction Strategy for Extended Hamming Codes in the EBD Framework}
\label{app:offline_details}
The iterative construction strategy in Steps 2 and 3 of the general EBD framework can be optimized. 
For convenience of description, let $Q_e$ and $Q_o$ denote the largest even integer and the largest odd integer less than or equal to $Q$, respectively. 
The following construction strategy ensures that all even-sized or odd-sized $\mathcal{Z}_t(\bm{s})$ with $t\leq Q$ can be obtained when $\bm{s} \in \mathcal{W}$ or $\bm{s} \in \mathcal{Y}$, respectively. 
Note that the process described as ``from $ a $ to $ b $'' can only be executed when $ a \leq b $.

\begin{itemize}

\item When $ \bm{s}\in\mathcal{W} $, let $ w $ be the even integer among $ Q_e/2 $ and $ Q_e/2 + 1 $.  
First, compute $ \mathcal{Z}_2(\bm{v}) $ for all $ \bm{v} \in {\mathcal{W}} $ with  $ t_1 = t_2 = 1 $.  
Then, for even $ t $ from 4 to $ w $, compute $ \mathcal{Z}_t(\bm{v}) $ for all $ \bm{v} \in {\mathcal{W}} $ using $ t_1=t_2=t/2 $ if $ t/2 $ is even, or $ (t_1, t_2) = (t-2, 2) $ otherwise.  
Finally, for even $ t $ from $ w+2 $ to $ Q_e $, compute $ \mathcal{Z}_t(\bm{s}) $ using $ t_1=t_2=t/2 $ if $ t/2 $ is even, or $ (t_1, t_2) = (t - w, w) $ otherwise.

\item When $ \bm{s}\in\mathcal{Y} $ and $ \lceil Q_o/2 \rceil$ is even, let $ w = \lceil Q_o/2 \rceil $.  
First, compute $ \mathcal{Z}_2(\bm{v}) $ for all $ \bm{v} \in {\mathcal{W}} $ with  $ t_1 = t_2 = 1 $.  
Next, for odd $ t $ from 3 to $ w-1$, compute $ \mathcal{Z}_t(\bm{v}) $  for all $ \bm{v} \in {\mathcal{Y}} $ using $ (t_1, t_2) = (t - 2, 2) $.  
Then, compute $ \mathcal{Z}_w(\bm{v}) $ for all $ \bm{v} \in {\mathcal{W}} $ using $ t_1 = t_2 = w/2 $ if $ w/2 = 2 $ or $ w/2 $ is odd, or $ (t_1, t_2) = (w-1, 1) $ otherwise (this step is performed only when $ w \neq 2 $). 
Finally, for odd $ t $ from $w+1$ to $ Q_o $, compute $ \mathcal{Z}_t(\bm{s}) $ using $ (t_1, t_2) = (t - w, w) $.

\item When $ \bm{s}\in\mathcal{Y} $ and $ \lceil Q_o/2 \rceil $ is odd, let $ w = \lceil Q_o/2 \rceil+1 $.  
First, compute $ \mathcal{Z}_2(\bm{v}) $ for all $ \bm{v} \in {\mathcal{W}} $ with  $ t_1 = t_2 = 1 $.  
Next, for odd $ t $ from 3 to $ w -3 $, compute $ \mathcal{Z}_t(\bm{v}) $ for all $ \bm{v} \in {\mathcal{Y}} $ using $ (t_1, t_2) = (t - 2, 2) $.  
Then, compute $ \mathcal{Z}_{w-1}(\bm{s}) $  using $ (t_1, t_2) = (w-3, 2) $, and compute $ \mathcal{Z}_w(\bm{v}) $ for all $ \bm{v} \in {\mathcal{W}} $ using $ t_1 = t_2 = w/2 $ if $ w/2 = 2 $ or $ w/2 $ is odd, or $ (t_1, t_2) = (w-3, 3) $ otherwise.  
Finally, for odd $ t $ from $w+1$ to $ Q_o $, compute $ \mathcal{Z}_t(\bm{s}) $ with  $ (t_1, t_2) = (t - w, w) $.

\end{itemize}

\begin{table}[htbp]
\centering
\caption{Specific Examples of Optimized Iterative Construction Strategy for Different Extended Hamming Codes}
\label{tab:iter}
\setlength{\tabcolsep}{8pt}  
\begin{tabular}{c ccc}
\toprule
Code    &  $Q$ & $\bm{s}\in{\mathcal{W}}$ & $\bm{s}\in\mathcal{Y}$ \\ 
\midrule
 $(32,26,4)$ &  6 & $[1,\tilde{2},\tilde{4}], (6)$  & $[1,\tilde{2},\tilde{4}], (3,5)$   \\
$(64,57,4)$ & 7 & $[1,\tilde{2},\tilde{4}], (6)$  & $[1,\tilde{2},3,\tilde{4}], (5, 7)$\\
$(128,120,4)$ &  8 & $[1,\tilde{2},\tilde{4}], (6,\tilde{8})$  & $[1,\tilde{2},3,\tilde{4}], (5, 7)$\\
$(256,247,4)$ &  9 & $[1,\tilde{2},\tilde{4}], (6,\tilde{8})$  & $[1,\tilde{2},3,\tilde{6}], (5, 7, 9)$\\
$(512,502,4)$ &  10 & $[1,\tilde{2},\tilde{4},6], (\tilde{8},10)$  & $[1,\tilde{2},3,\tilde{6}], (5, 7, 9)$\\

\bottomrule
\end{tabular}
\end{table}

\Cref{tab:iter} provides several specific  examples, where the last two columns describe how to construct error-building blocks when $\bm{s} \in \mathcal{W}$ and $\bm{s} \in \mathcal{Y}$, respectively. 
A number $t$ enclosed in square brackets (e.g., $[t]$) indicates that all non-offline-excludable target error-building blocks of size $t$ should be computed, while a number $t$ in parentheses (e.g., $(t)$) denotes that only the target $t$-error-building block for $ \bm{s} $ are required. A tilde above a number $t$ (e.g., $ \tilde{t} $) indicates that the current iteration can further halve the number of operations by setting $ t_1 = t_2 = t / 2 $. For example, when decoding the $(256, 247, 4)$ extended Hamming code with $\bm{s}\in\mathcal{Y}$, we first initialize $\mathcal{Z}_1(\bm{v})=\mathcal{O}_1(\bm{v})$ for all $\bm{v} \in \mathcal{Y}$, and then sequentially compute $\mathcal{Z}_2(\bm{v})$ for all $\bm{v} \in \mathcal{W}$ (symmetry can be exploited by setting $t_1=t_2=1$), $\mathcal{Z}_3(\bm{v})$ for all $\bm{v} \in \mathcal{Y}$, $\mathcal{Z}_5(\bm{s})$, $\mathcal{Z}_6(\bm{v})$ for all $\bm{v} \in \mathcal{W}$ (symmetry can be exploited by setting $t_1=t_2=3$), followed by $\mathcal{Z}_7(\bm{s})$ and $\mathcal{Z}_9(\bm{s})$. Finally, we select $\mathcal{O}(\bm{s})$ from among $\{\mathcal{Z}_1(\bm{s}), \mathcal{Z}_3(\bm{s}), \mathcal{Z}_5(\bm{s}), \mathcal{Z}_7(\bm{s}), \mathcal{Z}_9(\bm{s})\}$ (note that $\mathcal{Z}_2(\bm{s})=\mathcal{Z}_4(\bm{s})=\mathcal{Z}_6(\bm{s})=\mathcal{Z}_8(\bm{s})=\emptyset$).    

\begin{remark}
The above iterative construction strategy is applicable to any extended Hamming code of length $N \geq 8$, and has been verified to achieve the minimum computational cost for lengths $N \leq 256$ within the context of our EBD-OffOpt framework. 
However, it may not be optimal for longer codes. For example, when decoding the $(512,502,4)$ extended Hamming code with $\bm{s}\in\mathcal{W}$, this method yields the iteration strategy "$[1,\tilde{2},\tilde{4},6], (\tilde{8},10)$", whereas the strategy "$[1,\tilde{2},\tilde{4},\tilde{8}], (6,10)$" achieves lower computational cost.
\end{remark}

\begin{figure*}[htbp]
    \centering
    \includegraphics[width=0.865\linewidth]{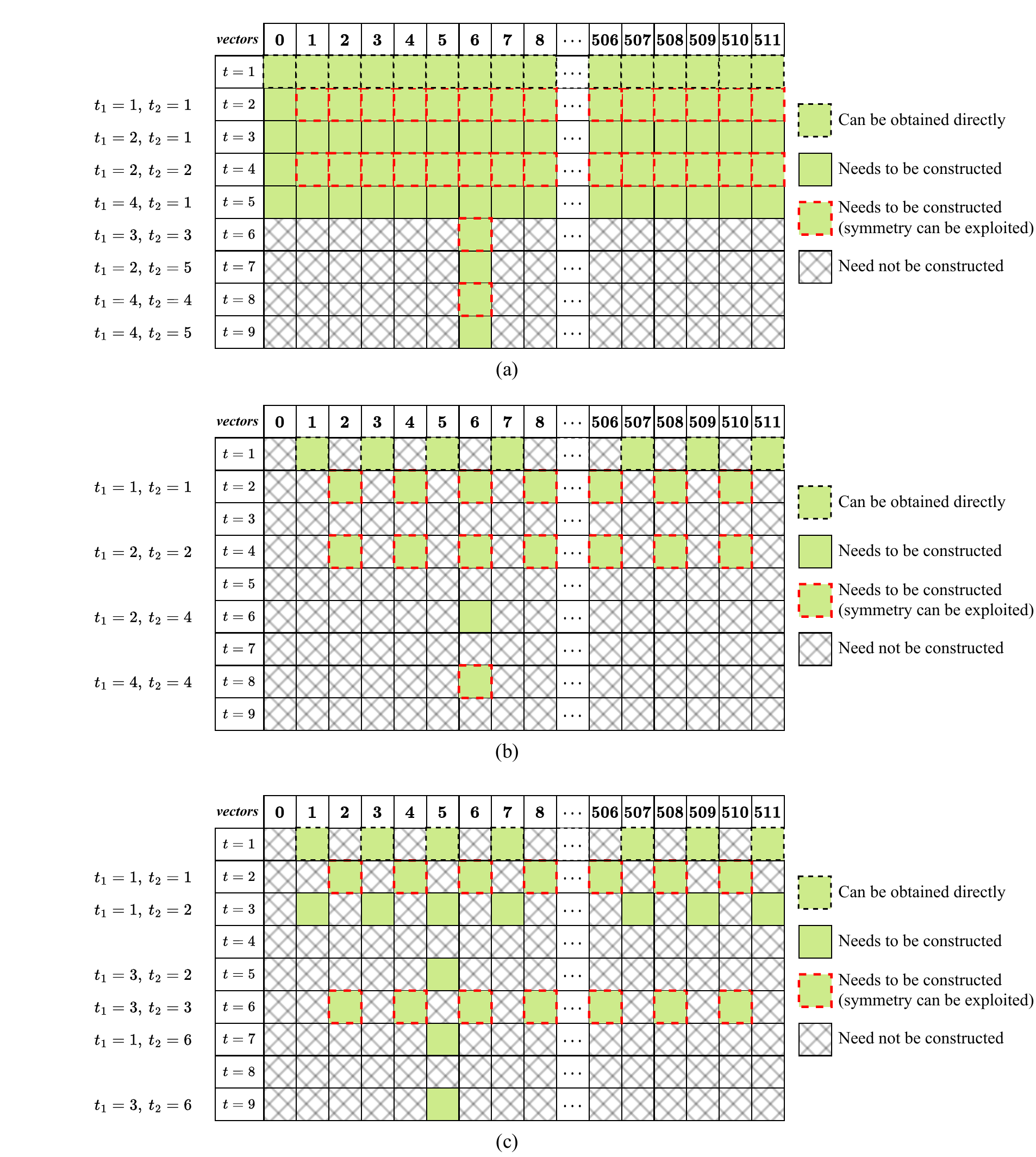} 
    \caption{Decoding of the $(256,247,4)$ extended Hamming code under: (a) the general EBD framework when $\bm{s}=\bm{6}$, (b) the optimized EBD-OffOpt framework (i.e., error-building decoding integrated with the offline exclusion mechanism) when $\bm{s}=\bm{6}$, and (c) the optimized EBD-OffOpt framework when $\bm{s}=\bm{5}$.} 
    \label{fig:ML_extendedhamming}
\end{figure*}

\Cref{fig:ML_extendedhamming} illustrates the iterative construction of error-building blocks for decoding the $(256, 247, 4)$ extended Hamming code under both the general and optimized EBD frameworks.
\Cref{fig:ML_extendedhamming} is read in a manner similar to \Cref{fig:ML_hamming}, but no concrete numerical examples are provided here, and the block penalty information is omitted. It can be observed that with the offline exclusion mechanism, the number of blocks requiring evaluation is significantly reduced, which means the construction process becomes much sparser and thus reduces decoding complexity. However, it should be noted that the construction process depicted in 
\Cref{fig:ML_extendedhamming}(a) is directly applicable to any other linear block code with $ Q = 9 $, as long as the parity-check matrix is known, and all constructed blocks (the green squares in this subfigure) are locally optimal; whereas \Cref{fig:ML_extendedhamming}(b) and (c) are customized versions  designed for the $(256, 247, 4)$ extended Hamming code, and the resulting blocks (the green squares in these subfigures) are not necessarily locally optimal.

\section{Details of Online Exclusion Mechanism}
\label{app:online_details}
Prior to constructing the target error-building blocks of size $t\geq2$ with the setting $t=t_1+t_2$, we need to obtain $\mathcal{D}_{t_j}^{t-1}(\bm{v}_i)$ for any $\bm{v}_i\in \mathbb{F}_2^Q$, where $t_j\in\{t_1,t_2\}$. While \Cref{equ:online_D} provides the original computational definition, we present here a more efficient approach for practical implementation, which is also employed in the experiments in \Cref{subsec:results}. 

Given $t_j$ and $\bm{v}_i$, if there exists some $t' \in [1, t-2]$ such that $\mathcal{D}_{t_j}^{t'}(\bm{v}_i)$ has already been computed at the time of constructing target error-building blocks of size $t'+1$, then the computation of $\mathcal{D}_{t_j}^{t-1}(\bm{v}_i)$ in \eqref{equ:online_D} is equivalent to 
\begin{equation}
\label{equ:online_D_alt}
    \mathcal{D}_{t_j}^{t-1}(\bm{v}_i) \!= \!
    \begin{cases}
    \mathcal{D}_{t_j}^{t'}(\bm{v}_i), 
    \!\!\!\!\! & \text{if } M\big(\mathcal{D}_{t_j}^{t'}(\bm{v}_i)\big) \!<\! M\big(\mathcal{Z}^{t-1}(\bm{s})\big), \\
    \emptyset,  \!\!\!\!\! & \text{if } M\big(\mathcal{D}_{t_j}^{t'}(\bm{v}_i)\big) \!\geq\! M\big(\mathcal{Z}^{t-1}(\bm{s})\big).
    \end{cases}
\end{equation}
This suggests a more efficient approach:  
If no previously computed $ \mathcal{D}_{t_j}^{t'}(\bm{v}_i) $ (with $ t' \in [1, t-2] $) exists, compute $ \mathcal{D}_{t_j}^{t-1}(\bm{v}_i) $ via \eqref{equ:online_D};  
otherwise, let $ t' $ be the largest such index and check whether $\mathcal{D}_{t_j}^{t'}(\bm{v}_i)=\emptyset$ or $ M\big(\mathcal{Z}^{t'}(\bm{s})\big) = M\big(\mathcal{Z}^{t-1}(\bm{s})\big) $. 
If so, reuse the previous result: 
$ \mathcal{D}_{t_j}^{t-1}(\bm{v}_i) = \mathcal{D}_{t_j}^{t'}(\bm{v}_i) $;
otherwise, compute $\mathcal{D}_{t_j}^{t-1}(\bm{v}_i)$ using \eqref{equ:online_D_alt}.
Clearly, this approach can be implemented by performing an in-place update over the blocks of sizes $ t_j $ that have already been computed. 
Blocks set to empty during the update process indicate that they have been excluded online.


\end{document}